\documentclass[review]{elsarticle}
\usepackage{longtable}
\usepackage{lineno,hyperref}
\modulolinenumbers[5]
\usepackage[]{subfloat}
\journal{Journal of \LaTeX\ Templates}
\usepackage{graphicx}
\usepackage{lscape}
\usepackage{subfigure}
\usepackage{xcolor}
\usepackage{caption}
\usepackage{csquotes}
\usepackage{pgfplots}

\usetikzlibrary{fadings}

\begin{document}
	\begin{frontmatter}
		
		\title{A systematic literature review on machine learning applications for consumer sentiment analysis using online reviews}
		
		\author{Praphula Kumar Jain$^{1,*}$ and Rajendra Pamula$^1$}
		\address{$^1$ Indian Institute of Technology (Indian School of Mines), Dhanbad-826004, JH, INDIA}
		
		
		\cortext[mycorrespondingauthor]{Corresponding author}
		\ead{praphulajn1@gmail.com}
		

\begin{abstract}
Consumer sentiment analysis is a recent fad for social media related applications such as healthcare, crime, finance, travel, and academics. Disentangling consumer perception to gain insight into the desired objective and reviews is significant. With the advancement of technology, a massive amount of social web-data increasing in terms of volume, subjectivity, and heterogeneity, becomes challenging to process it manually. Machine learning techniques have been utilized to handle this difficulty in real-life applications. This paper presents the study to find out the usefulness, scope, and applicability of this alliance of Machine Learning techniques for consumer sentiment analysis on online reviews in the domain of hospitality and tourism. We have shown a systematic literature review to compare, analyze, explore, and understand the attempts and direction in a proper way to find research gaps to illustrating the future scope of this pairing. This work is contributing to the extant literature in two ways; firstly, the primary objective is to read and analyze the use of machine learning techniques for consumer sentiment analysis on online reviews in the domain of hospitality and tourism. Secondly, in this work, we presented a systematic approach to identify, collect observational evidence, results from the analysis, and assimilate observations of all related high-quality research to address particular research queries referring to the described research area. 

\end{abstract}

%


%
%
%
%
%
%
%
%
%
%
%
%
%
\end{frontmatter}
\tableofcontents
\newpage
\section{Introduction}
\label{intro}
With the enhancement of information and communication technology (ICT), increasing the volume of consumer feedback data in terms of online reviews has becomes a recent research topic in the field of Consumer Sentiment Analysis (CSA) \cite{107,108}. Online review based CSA considered as a trend for many real-life applications which includes behaviour analysis, decision making, and getting insight information for organizational growth. Interestingly, the online reviews, which are probably found in the unstructured form \cite{109} further, difficult to getting insight meaningful information manually. Natural Language Processing (NLP) and Text Mining (TM) models describe the transformational process and convert this unstructured data into structured one for Data Mining (DM). The use of Machine Learning (ML) techniques to brilliantly mine online reviews has been found broadly in literature \cite{110,111}. CSA, traditionally as a commonly DM and text classification task \cite{114}, is described as the computational understanding of consumer’s sentiments, opinions and attitude towards services or products \cite{112,113}. CSA provides a technological solution to sensing and understanding consumer experiences, views, and approach in online reviews available over the online platform.     

In recent literature, many researchers focused their study in CSA using online reviews in many domains such as healthcare \cite{115}, business \cite{116}, tourism and hospitality \cite{117}, academics \cite{118}, etc. Hence, CSA using online reviews has escalated in the last few years with the enhancement of ICT. This Improved ICT providing facility to write reviews, comments, feedback, or blogs, over the various online platforms such as Twitter, Facebook, Skytrax, Yelp, and Tripadvisor, etc. These online platforms are also open and freely available to post reviewers' experience regarding used services or products. These consumer provided reviews is helpful for both, service provider organizations and forthcoming consumer, organizations can use it in improving service quality and in making better consumer policies, and the forthcoming consumer can refer if before making their purchase decision. Hence, CSA has been introduced as an efficient way to extract the information expressed in the online reviews, especially in the context of hospitality and tourism. CSA extracts the consumer sentiment, opinion, and demands from the online reviews in a particular domain and identifies their polarity \cite{119}.  Nowadays, the exchange of consumer feeling through the online platform is the main objective of the researcher in CSA. 

Nowadays, if any consumer wants to buy a product or services, there is no need to ask their friends and family for previous consumers' experience because of the availability of many user reviews and discussion platforms regarding any product or services.  For any service provider organizations, there is no necessity to conduct consumer surveys, opinion polls, and group discussions in order to collect consumer experiences because of the abundance availability of such information in the form of online reviews over many websites. 
In the current scenario, we have found that providing online reviews publicly over the social media platform has helped and reshaped in business growth. 

\begin{figure}[h]
	\centering
	\includegraphics[width=12cm, height=7cm]{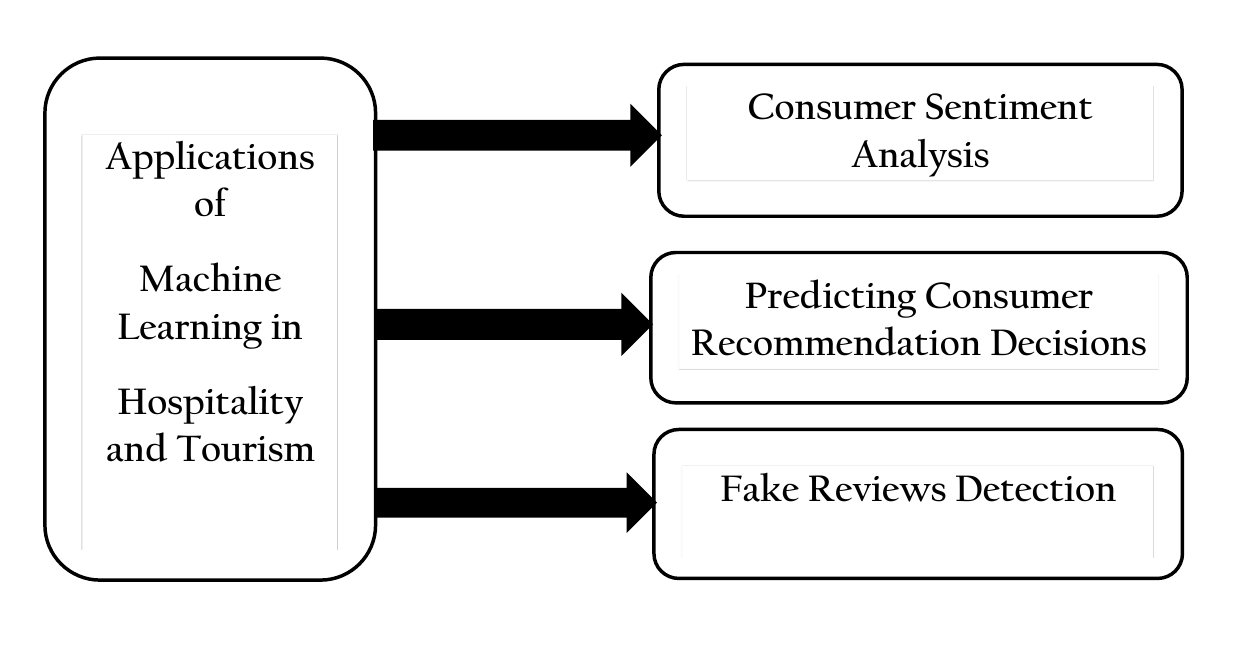}
	\caption{CSA using ML in the domain of Hosptality and tourism}
	\label{fig:11}       
\end{figure}

In recent research, few review studies were found on ML applications for sentiment analysis, these literature surveys have focused on ML applications in various domains such as twitter sentiment \cite{120}, scientific citations\cite{121}, business behaviour\cite{122}, reputation evaluation \cite{123} etc. However, to the best of our knowledge, no such literature survey is conducted to review the current status of ML applications in CSA using online reviews in the domain of hospitality and tourism. The increasing amount of data available over the online platform offers the CSAs new possibilities to predict consumer sentiment and use it in business growth. The practitioners should be equipped with the recent information to confirm that their drawings are significant form online reviews.   In this review paper, we included 68 articles on ML application in CSA, specifically in the domain of hospitality and tourism. We considered articles related to ML applications using online reviews in sentiment classification, predictive recommendation decision, fake reviews detection (Refer to Fig.\ref{fig:11} ). It is foreseen that in the future, the hospitality and tourism sector will continue to discover the many ML-based research articles and the observations provided in this study will help the researcher in sensing and understanding the current role of  ML applications in CSA, which will direct them to apply ML to support their CSA-related findings.

The rest of the paper is structured as follows: For the readers' comfortability, abbreviations utilized in this SLR are mentioned in Table \ref{abb}. Section 2 describes a brief introduction to CSA and ML algorithms. Section 3 presents the SLR process implemented in this paper. Section 4 presents the findings of the SLR. An ML-CSA framework and implications based on the results of the review are discussed in Section 5. The conclusions and limitations of the study are described in details in Section 6.

\begin{center}
	\begin{longtable}{ll}
		\caption{Abbreviations} \label{abb} \\
		\hline \multicolumn{1}{c}{\textbf{Abbreviation}} & 
		\multicolumn{1}{c}{\textbf{Description}}\\ \hline
		\endfirsthead
		\multicolumn{2}{c}%
		{{\bfseries \tablename\ \thetable{} -- continued from previous page}} \\
		\hline \multicolumn{1}{c}{\textbf{Abbreviation}} & 
		\multicolumn{1}{c}{\textbf{Description}}\\ \hline 
		\endhead
		\hline \multicolumn{2}{r}{{Continued on next page}} \\ \hline
		\endfoot
		\hline \hline
		\endlastfoot
		
		\noalign{\smallskip}\hline\noalign{\smallskip}
		ML& Machine Learning\\
		NB& Naive Base\\
		LR& Logistic Regression\\
		SVM& Support Vector Machine\\
		RF& Random Forest\\
		DT& Decision Tree\\
		NN&Neural Network\\
		ANN& Artificial Neural Network\\
		KNN&K-Nearest Neighbors\\
		PCA&Principal Component Analysis\\
		GB&Gradient Boosting\\
		LSA&Latent Semantic Analysis\\
		ANFIS&Adaptive Neuro-Fuzzy Inference System\\
		CTK&Convolution Tree Kernel\\
		CNN&Convolutional Neural Network\\
		ADLM&Autoregressive Distributed Lag Model\\
		RBFNN&Radial Basis Function Neural Network \\
		MLP & Multi-Layer Perceptron\\
		SVR & Support Vector Regression\\
		BPNN & Backpropagation NN\\
		TFM & Transfer Function Model\\
		ARIMA & AutoRegressive Integrated Moving Average\\
		AR-NN& Autoregressive Neural Network\\
		LSTM& Long Short-term Memory\\
		BiLSTM&Bidirectional LSTM\\
		BBiLSTM& Bayesian Optimization BiLSTM  \\
		ADLM&  Autoregressive Distributed Lag Model\\
		GP&Gaussian Processes\\
		DSA&Data-based Sensitivity Analysis\\
		$x_{t}$&Actual\\
		$\hat{x_{t}}$&Predicted\\
		
		MdAPE&Median Absolute Percentage Error\\
		RMSE&Root Mean Square Error\\
		MAE&Mean Absolute Error\\
		MAPE&Mean Absolute Percentage Error \\
		RRMSE&Ratio of RMSE \\
		
		TP&True Positive\\
		TN&True Negative\\
		FP&False Positive\\
		FN&False Negative\\
		\noalign{\smallskip}\hline
	\end{longtable}
\end{center}
\section{CSA: Process, technniques, Challanges}
\subsection{Consumer sentiment analysis (CSA) using online reviews}
\label{2.1}
CSA using online reviews is a process of extracting whether a part of online reviews (product/services reviews, tweets, online surveys, etc.) is negative, positive, or neutral. CSA may be used to identify the consumer's attitude regarding service by the use of words such as opinion, tone, context, etc. Organizations may use CSA to collect previous consumer experiences of their services or products and may use their findings in service improvement \cite{132}. Organizations may also utilize this analysis to collect valuable consumers' experience about the issues in newly released products. CSA not only helps organizations to understand consumer satisfaction regarding their offerings, but it also provides a better understanding of how they lack form their competitors. CSA is also helpful for new consumers to gain knowledge about products or services before making purchase decisions. 

\begin{figure}[h]
	\includegraphics[width=12cm, height=8cm]{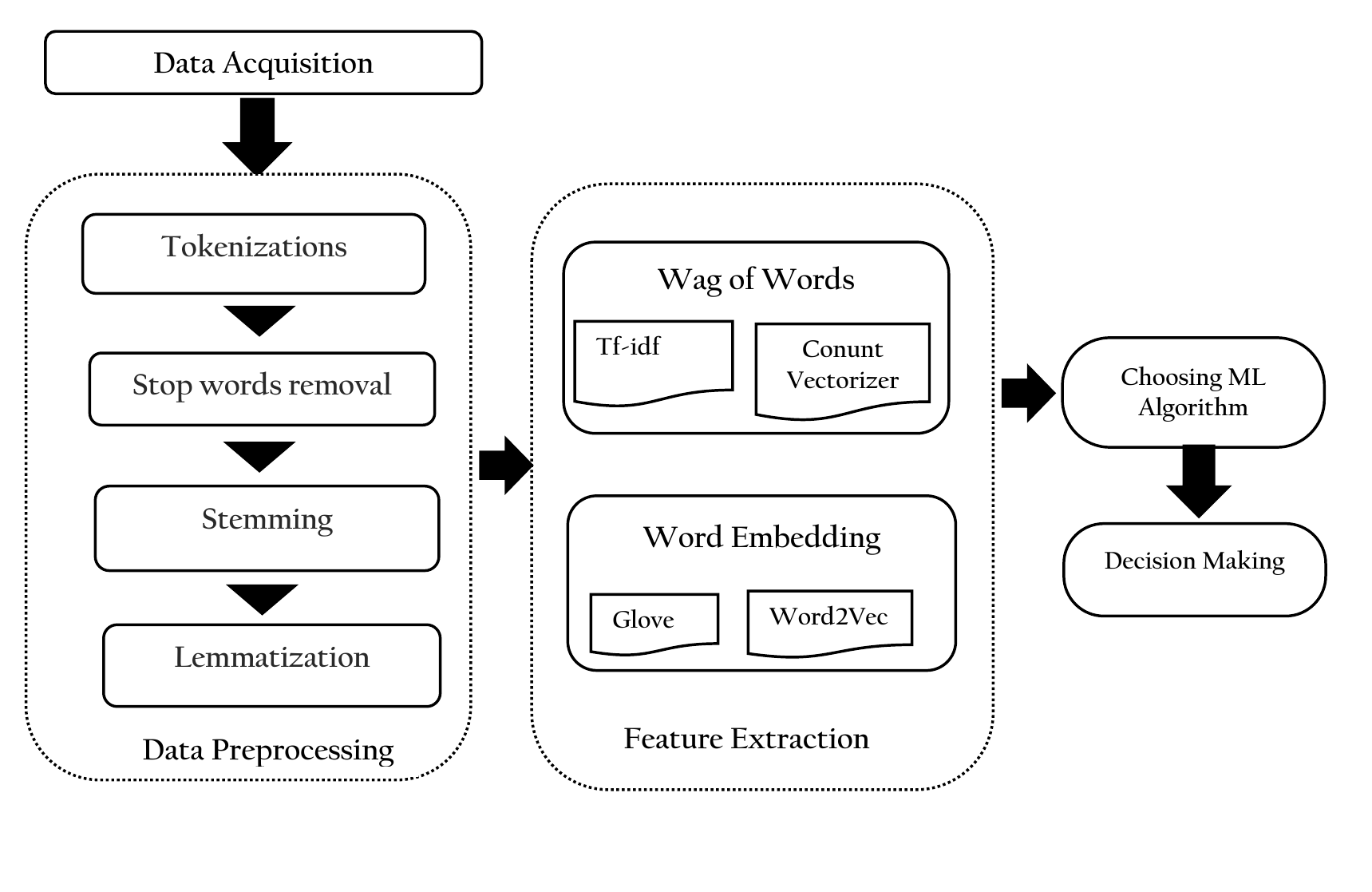}
	\caption{CSA process diagram}
	\label{fig:CSA process diagram}       
\end{figure}

A typical CSA process diagram is presented in Fig \ref{fig:CSA process diagram}, data processing steps with descriptions are mentioned in Table \ref{tab:33}, and various feature selection techniques with explanations are mentioned in the Table \ref{tab:22}.

\begin{table}[h]
	\caption{Text feature selection techniques with descriptions}
	\label{tab:22}       
	\begin{tabular}{p{0.25\textwidth}p{0.2\textwidth}p{0.6\textwidth}}
		\hline\noalign{\smallskip}
		& Technique  & Description  \\
		\noalign{\smallskip}\hline\noalign{\smallskip}
		Wag of Words& Tf-idf&  Tf-idf stands for term frequency-inverse document frequency, and the tf-idf weight is a weight often used in information retrieval and text mining. This weight is a statistical measure used to evaluate how important a word is to a document in a collection or corpus \cite{187}.  \\
		& CountVectorizer& The CountVectorizer offers a simple way for both tokenizations of collection of text documents and creating a vocabulary for established words, but also for encoding new documents using that vocabulary \cite{188}.  \\
	Word Embeddings&Word2Vec& This is a two-layer neural network that processes text by "vectorizing" words, where input is a text corpus, and output is a set of vectors: feature vectors representing words in that corpus. It converts text into a numerical form that neural networks can process \cite{189}.  \\
	& Glove&GloVe (Global Vectors for Word Representation) is used to get word representations of vectors. Training is carried out from a corpus on aggregated global word-word co-occurrence statistics, and the resulting observations reveal fascinating linear substructures of the word vector space \cite{190}.\\
		\noalign{\smallskip}\hline
	\end{tabular}
\end{table}

\begin{table}[h]
	\caption{Text preprocessing steps with description}
	\label{tab:33}       
	\begin{tabular}{p{0.25\textwidth}p{0.6\textwidth}}
		\hline\noalign{\smallskip}
		Step  & Description  \\
		\noalign{\smallskip}\hline\noalign{\smallskip}
	Tokenizations& Tokenization is a way of dividing textual statements into words called tokens \cite{192}.   \\
		Stop words removal& Stop word are the words that not have significance in search queries, and process of removing stop words (is, am, a, an, etc.) called as stop worrd removal \cite{193}.  \\
		Stemming& Process of reducing a word to its word stem that affixes to suffixes and prefixes or to the roots of words known as a lemma \cite{191}.  \\
		Lemmatization&  Lemmatization, unlike Stemming, reduces the inflected words properly ensuring that the root word belongs to the language \cite{191}. \\
		\noalign{\smallskip}\hline
	\end{tabular}
\end{table}
\subsection{Challenges for CSA}
\label{2.2}
Extracting and collecting consumer sentiment form online reviews is a part of information fusion or information extraction \cite{132} whose objective is to identify the context of the used words. This needs a huge amount of data set to train the model and also required particular domain knowledge about the objective of the study.  With the competitive era, there is also a possibility of fake reviews so before making consumer decision based on the online reviews need to be careful. For making the significant consumer sentiment decisions, it is difficult to discard the fake reviews, and unopinion content as these may confuse the proposed model while classifying sentiment. In this subsection, we are summarizing the CSA challenges.\\

\noindent\textbf{\emph{Subjectivity Detection:}} The very first difficulty is that subjectivity identification itself is a subjective task, i.e., a part of the online reviews may be a neutral to a few consumers but not to others. This can be seen because of the heterogeneity of the consumers on a single review subject but also because of various interpretations of a sentence in different languages. In some of the context, this may be related to the preferences by a particular one by their personality. In this context, before subjectivity detections, there is a need to perform user profiling such as personality detection \cite{135}, gender identification \cite{133}, and community detection \cite{134} in advance. \\

\noindent\textbf{\emph{Accuracy improvement of subjectivity Detection:}}
The next difficulty is the accuracy improvement of the subjectivity detection task in a short text. In some of the cases, collected data may have short sentences, and they may not have sufficient contextual information, so it is difficult to classify the sentences based on the short sentences. In some instances, due to missing details in the short text, regularisation is required. In these situations, false reviews are highly likely because of the lack of characteristics that describe good or negative behaviour. To this end, we need to build extra data samples using the standardization techniques, using generative adversarial networks.\\

\noindent\textbf{\emph{Context dependency:}} The next difficulty is context dependency. There are few words present in the text that are objective, but in some cases, they are subjective e.g., the objective ‘long’ is considered a positive in some domains, e.g., ‘long battery,’ or considered as negative in some cases, e.g. ‘long queue,’ actually word long do not show any behavior as positive or negative. The meaning of a word is often based on terms far away in the sentence and often beyond the neighboring word window. To identify the substructure of the word, parse tree models is essential in conjunction with word vectors. \\

\noindent\textbf{\emph{Computational cost:}}  
The last challenges are computational cost reduction of the training attributes because it is difficult and time consuming to process large vocabulary of words. With the help of traditional methods, feature identification, and subjectivity detection of sentences is difficult, so in this SLR, we have considered ML techniques in processing online reviews content. 
\subsection{Machine learning techniques}
ML described as "the experimental study of algorithm and computational models over computer utilizing the previous experience for continuous performance improvements on a particular task or making prediction accurately" \cite{124}. The word "experience" in the mentioned description refers to the previous data available to the practitioner for making a predictive or classification model. These datasets may be gathered from an online data source or from the conducting survey over the online or offline platform. ML related comman terminology listed in Table \ref{tab:11}.  

\begin{figure}[h]
 	\includegraphics[width=12cm, height=8cm]{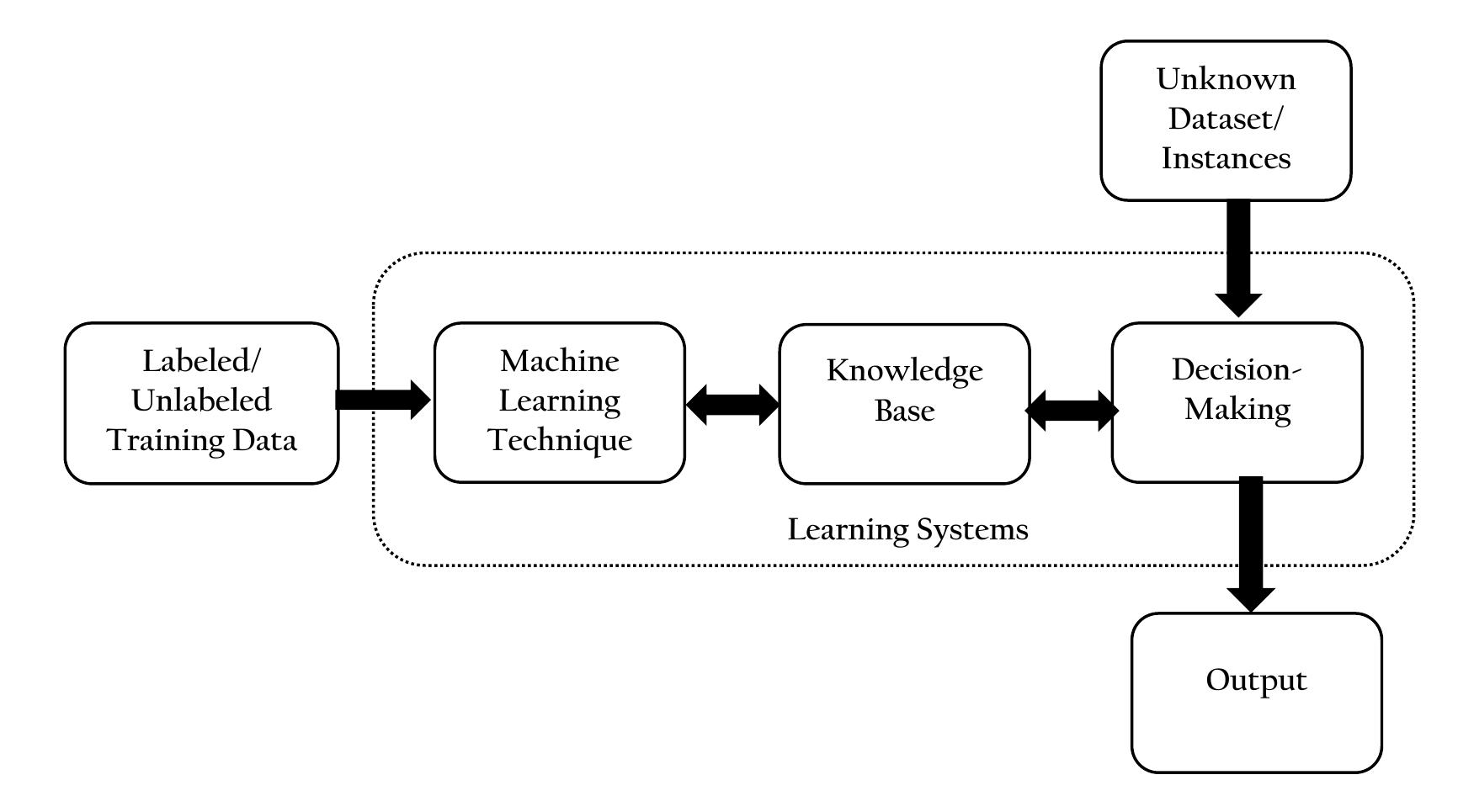}
 	\caption{A machine learning system configuration (taken from \cite{104} ).}
 	\label{fig:2}       
 \end{figure}

\begin{table}[h]
	\caption{Machine learning (ML) Technologies (refered from \cite{124})}
	\label{tab:11}       
	\begin{tabular}{p{0.25\textwidth}p{0.7\textwidth}}
		\hline\noalign{\smallskip}
		Word  & Description  \\
		\noalign{\smallskip}\hline\noalign{\smallskip}
		Example& Data instances used for learning   \\
		Feature& Collection of attribute(s) or vectors related with an instance.  \\
		Label& Value(s) assigned to the examples.\\
		Test sample&Examples utilized for evaluating the performance of a learning algorithm.\\
		Training Sample&Examples utilized for training a learning algorithm.\\
		Validation Sample&Examples  utilized for tuning the parameters of a
		learning sample.\\
		\noalign{\smallskip}\hline
	\end{tabular}
\end{table}
A customarily configured ML architecture shown in Fig \ref{fig:2}. For the ML systems, the input is the labelled or unlabelled training data gathered from various sources. The knowledge base is useful in deciding on appropriate ML technique for a particular task. In the ML system, previous findings or information are also helpful in validating the ML classification or predictions gained from the recent data for the accurate output and enhancing the decision making ability of the algorithms \cite{125}.

The ML algorithms have been divided into three major categories; supervised learning, unsupervised learning, and reinforcement learning. In the case of supervised learning, we have prior knowledge of input and required output, classification or prediction model developed based on the labelled dataset.  The main objective of the supervised learning algorithms is to obtain the desired output based on the available inputs \cite{126,127}. Decision Tree (DT), Random Forst (RF), and regression analysis are categorized into the supervised learning algorithms. In unsupervised learning, classification or predictive model builds without any previous knowledge of input and output variables. The primary task of unsupervised learning is dimensionality reduction and exploratory analysis \cite{128}, includes Neural Network (NN), Deep Learning (DL), Clustering, and optimization techniques. In reinforcement learning \cite{129,130}, there is not any presence of supervisor and labelled dataset, the training and testing datasets are mixed, and the learner collects information by interacting with the environment. Deep-Q-learning, Q-learning, and robot navigation techniques come under reinforcement learning.   
  
\begin{figure}
		\includegraphics[width=13cm, height=8cm]{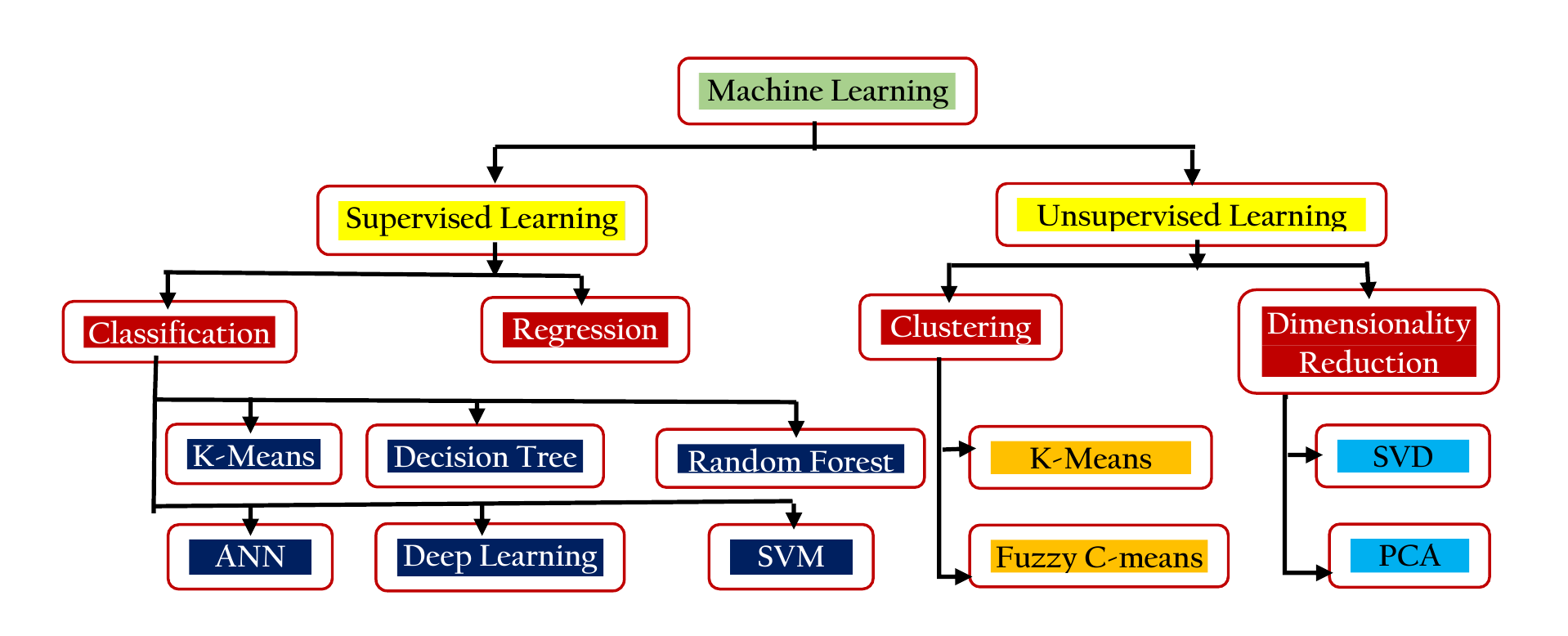}
		\caption{ML techniques classification (refered from \cite{131})}
		\label{fig:classification}       
\end{figure}

A short description of ML algorithms used in CSA shown in Table \ref{tab:2} and their classification is described in Fig \ref{fig:classification}.

\begin{center}
\begin{longtable}{p{0.19\textwidth}p{0.8\textwidth}}
		\caption{Description of Machine learning (ML) Algorithm }
	\label{tab:2}\\
	\hline \multicolumn{1}{c}{\textbf{ML Algorithm}} & \multicolumn{1}{c}{\textbf{Description}}\\ \hline
	\endfirsthead
	\multicolumn{2}{c}%
	{{\bfseries \tablename\ \thetable{} -- continued from previous page}} \\
	\hline \multicolumn{1}{c}{\textbf{ML Algorithm}} & \multicolumn{1}{c}{\textbf{Description}}\\ \hline
	\endhead
	\hline \multicolumn{2}{r}{{Continued on next page}} \\ \hline
	\endfoot
	\hline \hline
	\endlastfoot
\noalign{\smallskip}\hline\noalign{\smallskip}
		DT&  This technique classifies the data into smaller subgroups in which each subgroup includes (mostly) one-class responses, either "yes" or "no."  \cite{173,174}\\
		Regression & This is a classical predictive model that represents the relationship between inputs and output variables in the form of an equation \cite{175,176}. There are many forms of regression studied in literature few of them are linear regression, ridge regression, logistic regression, lasso regression, and polynomial regression, etc. \\
		SVM&This is a classification technique that determines multidimensional boundaries separating data points belonging to various classes  \cite{177}.\\
		Clustering & Clustering techniques like k-means divided the data points into k-clusters and by calculating k centroids \cite{178}.\\
		NN&This is a computational and mathematical model influenced by the biological neurons, in NN weight adjusted to reduce error between real and predicted  \cite{179,180}. NNs are further divided into multilayer perceptron, backpropagation NN, and Hopfield networks.\\
		NB&This is a classification technique that predicts the output class by using  Bayes’ theorem based on calculating conditional probability and prior probability \cite{181}.\\
		RF&A classifier in which "forest" is built based on the ensemble of decision trees and typically trained with the process of "bagging"\cite{182}.\\
		LSTM&This is a special type of recurrent neural network (RNN) that is capable of learning long term dependencies in data \cite{183}. \\
		BiLSTM&This is a sequence processing model that consists of two LSTMs: one is taking input in a forward direction, and the other in a backwards direction \cite{184}.\\
		CNN& This is a deep neural network used in ML and deep learning \cite{185}.\\
		LSA&This is a fundamental topic Modeling technique, in which words that are similar in meaning come in the same excerpt of the text \cite{186}.\\
		\noalign{\smallskip}\hline
		\end{longtable}
\end{center}

\section{Systematic Literature Review (SLR) Methodology}
\begin{figure}
	\centering
	\includegraphics[width=13cm, height=8cm]{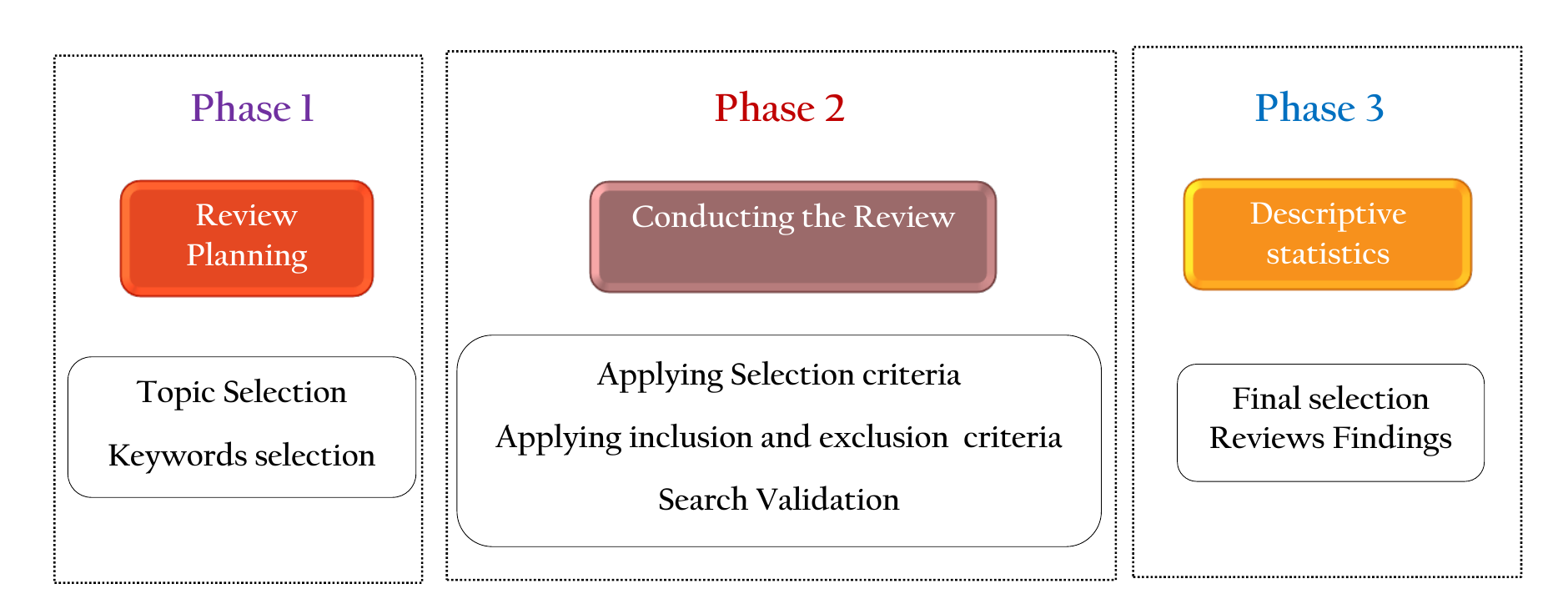}
	\caption{Three Phase SLR applied in CSA }
	\label{fig:14}       
\end{figure}

In this work, we used a systematic literature review (SLR) with a particular objective on surveying the published research article accurately and maintaining unbiasedness and purpose summary of the recent research era and future direction of ML applications in CSA. SLR influenced by a scientific methodology and motivated \cite{100} for explaining and evaluating all the present research related to the query, topic, or paradox of passion \cite{101}. SLRs can be a valuable contribution in drawing useful insights form the theoretical analysis of existing literature and finding possible flaws in the available research \cite{102}. In this survey, we applied a three-stage SLR process (refer to Fig. \ref{fig:14}) advised by the authors in \cite{103}, constituting the review planning, conducting the review, and reviews findings.     

\subsection{Review Planning (pre-operational stage)}
The goal of this review was to study the applications of ML techniques in CSA. The focus was to study ML techniques, as mentioned in Table \ref{tab:2}, on the CSA process depicted in Fig. \ref{fig:CSA process diagram}. More specifically, we were sensing and understanding how the utilization of ML techniques helpful in making the CSA process efficiently. To restrict the conceptual boundaries, we utilize preferred keywords covering ML techniques such as Naive base (NB), Support Vector Machine (SVM), Neural Network (NN), and regression, etc. Not many restrictions were implemented to the keywords for “consumer sentiment analysis” and search applied based on broad terms such as "consumer sentiment",  "consumer opinion", and "consumer views". All the selected research articles reviewed to test our objective of review and to evaluate whether CSA  considered in the scope of the article or not.

\subsection{Conducting the Review (operational stage)}
We have initially searched CSA related articles over the online platform www.scholar.google.com,  based on the search keywords formed in the first stage.  This website database was selected because it consists of a variety of refereed journals from reputed publishers such as IEEE, Springer, Elsevier, Taylor and Francis, IET, and SAGE, etc. Later the search also carried on the Scopus database so that we can include more and more related articles to CSA. However, to restrict the number of articles for review, we adopted inclusion and exclusion criteria as mentioned:
\subsubsection{Inclusion Criteria}
\begin{itemize}
	\item Articles published in SCI/SSCI/SCIE indexed journals or Q1/Q2 journals in hospitality and tourism. 
	\item Articles published From January 2017 to July 2020, related to CSA based on online reviews specifically to hospitality and tourism.
	\item Articles related to CSA based on online reviews specifically to the hospitality and tourism. 
	\item Articles are demonstrating CSA using ML techniques in hospitality and tourism. 
	\item Articles are demonstrating CSA using Hybrid ML techniques in hospitality and tourism.
	\item Articles are demonstrating CSA using Deep Learning (DL techniques in hospitality and tourism.
	\item Articles are demonstrating the comparison of ML techniques in CSA in hospitality and tourism. 
	\item Articles are demonstrating CSA of online reviews written in the English language only.
	
\end{itemize}
\subsubsection{Exclusion Criteria}
\begin{itemize}
	\item Articles published in conferences. 
	Articles published in Book Chapter/trade journals/book contributions.  
	\item Articles that are without justifiable research contribution or benchmark comparisons.
	\item Articles using any other online reviews data like a business, healthcare, or academics, etc
	\item Articles with qualitative and quantitative data considered, other forms of data like image, video, and audio not considered. 
	\item Articles that are reviews or surveys on CSA without any findings.
	\item Articles involving CSA on languages other than English, for example, languages Hindi, Portuguese, Chinese, Bengali, Arab, Spanish, etc. are not considered.
	\item Articles published before January 2017. 
	
\end{itemize}
We have applied a Backward and forward search to confirm that the selected articles are extensive and significant research in CSA deploying ML techniques are included.  A total of 253 articles were selected as an outcome by utilizing the mentioned keywords; machine learning or consumer sentiment analysis or consumer views or recommendation prediction or fake reviews or decision tree or support vector machine or consumer intention or neural network. The search based on keywords was applied based on a pairwise combination form ML techniques and others from its application into hospitality and tourism.  

\begin{figure}[h]
	\centering
	\includegraphics[width=13cm, height=8cm]{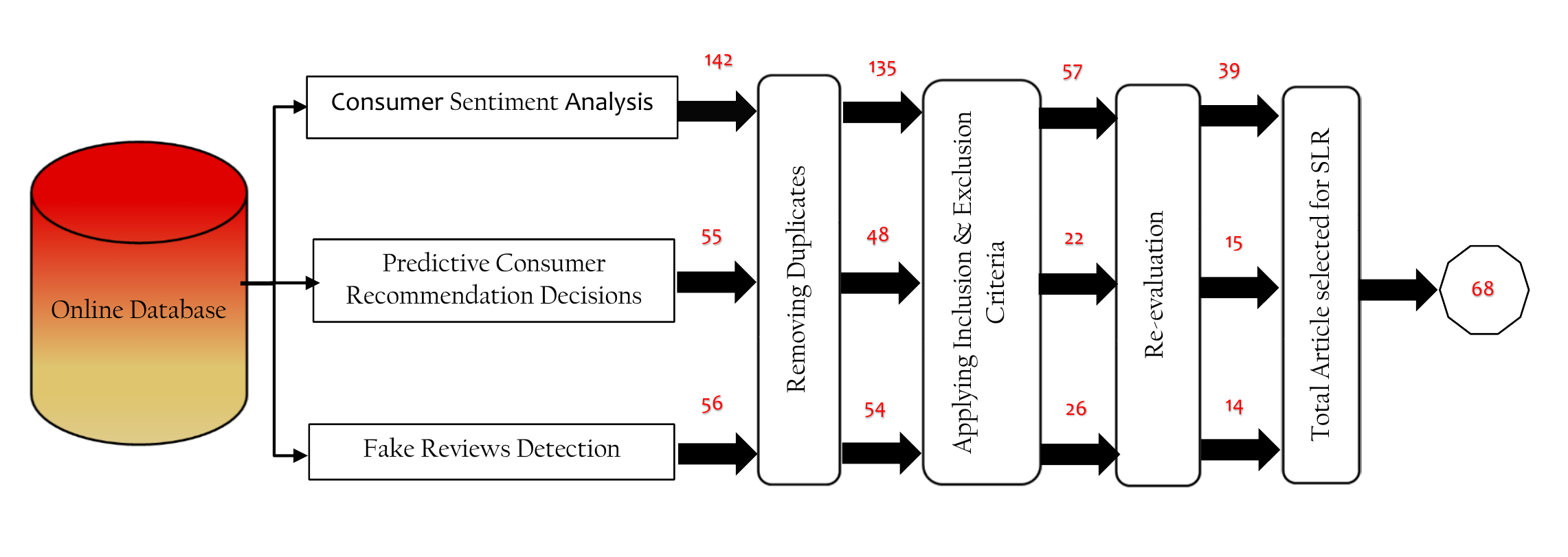}
	\caption{Publication selection process for SLR}
	\label{Fig:a}       
\end{figure}

The selected articles were validated before making the final selection for review. All the authors re-evaluated all chosen articles, and understanding was coded to form reliability. The same articles selected by all the authors were selected for the study, and rest articles re-evaluated. After re-evaluation of the articles' final selection was made for the review process, any articles that do not find suitable for review excluded from the study \cite{105}.  The validation stage helps in including a quality article in a related area to the objective of the review, instead of believing in the journal quality rating. The mentioned selection process derived in a comprehensive 68 articles since it concludes the quality of articles published to CSA in the hospitality and tourism domain.  The article selection process is shown in Fig. \ref{Fig:a}

\subsection{Descriptive statistics}
 
\begin{figure}
	\centering
	\includegraphics[width=12cm, height=10cm]{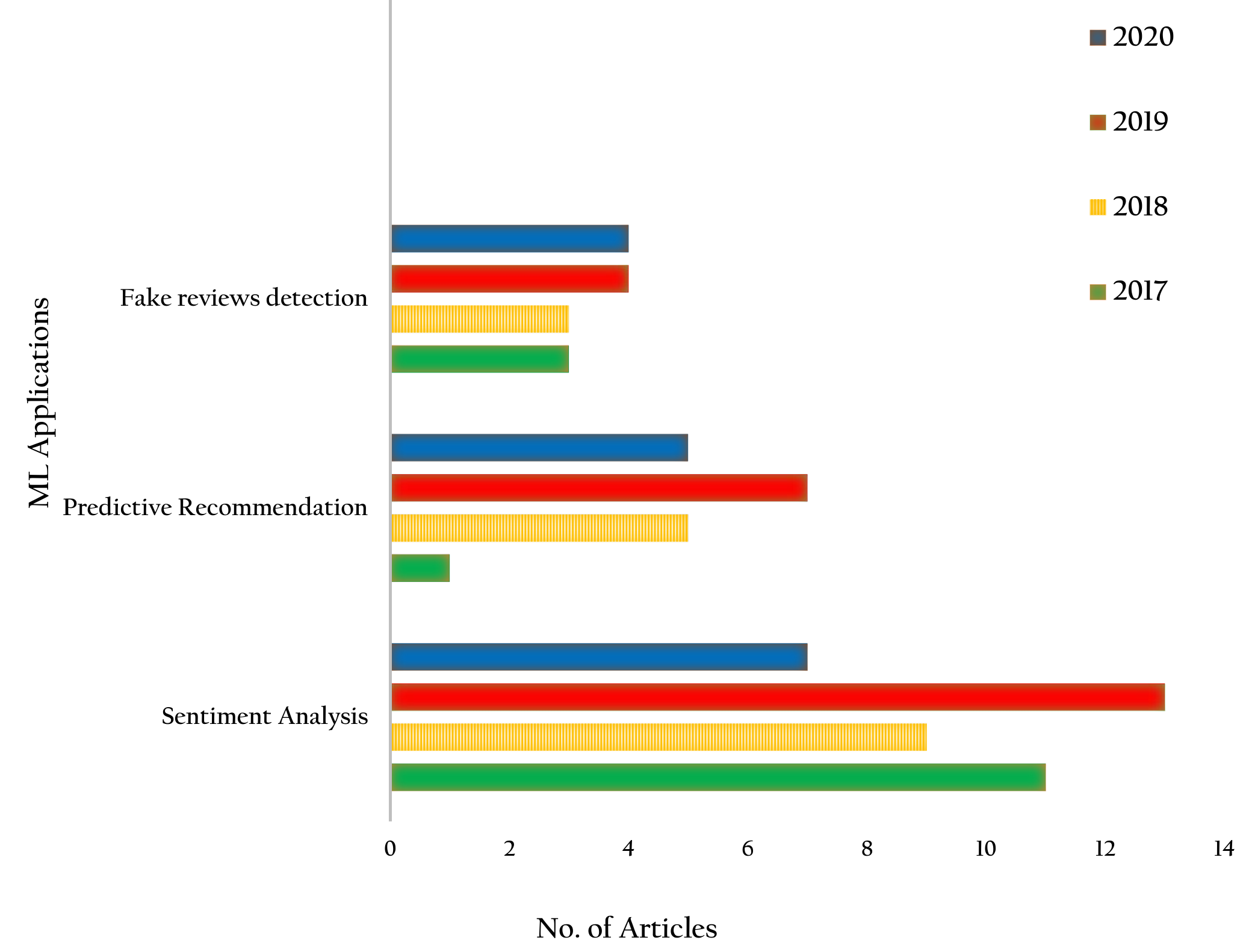}
	\caption{Year wise article published on different sub-area of CSA}
	\label{fig:65}       
\end{figure}

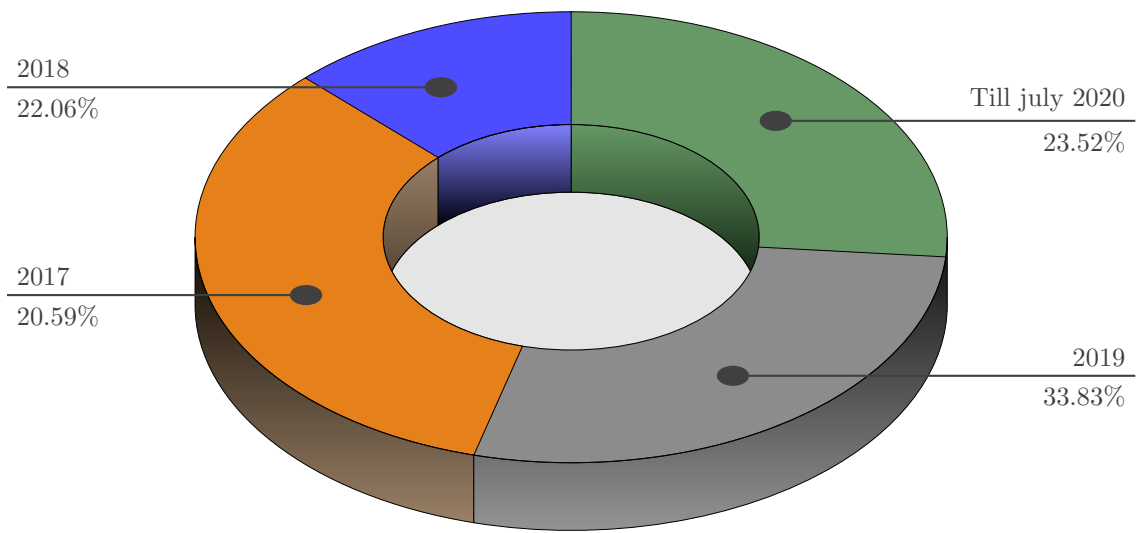
\begin{figure}
	\begin{tikzpicture}[fading style/.style={preaction={fill=#1,opacity=.8,
			path fading=circle with fuzzy edge 20 percent}}]
	
	\begin{scope}[xscale=5,yscale=3]
	\path[fading style=black,transform canvas={yshift=-40pt}] (0,0) circle (1cm);         
	\fill[gray](0,0) circle (0.5cm);  
	\path[fading style=white,transform canvas={yshift=-16mm}] (0,0) circle (0.65cm);    
	\draw[yshift=-3mm](0,0) circle (0.5cm); 
	
	\shadedraw[top color=green!20!gray,,bottom color=green!5!black,draw=black,very thin]  
	(90:0.5cm)--++(0,-3mm) arc(90:-5:0.5cm)--++(0,3mm)  arc(-5:90:0.5cm)--cycle;      
	\shadedraw[top color=orange!20!gray,bottom color=orange!5!black,draw=black,very thin]   
	(-105:0.5cm)--++(0,-3mm) arc(-105:-225 :0.5cm)--++(0,3mm)  arc(-225:-105:0.5cm)--cycle;   
	\shadedraw[top color=blue!50!white,,bottom color=blue!5!black,draw=black,very thin]   
	(135 :0.5cm)--++(0,-3mm) arc(135:90:0.5cm)--++(0,3mm)  arc(90:135:0.5cm)--cycle;  
	
	\begin{scope}[draw=black,thin]
	\fill[green!20!gray](90 :0.5cm)--(90:1cm) arc(90:-5:1cm)--(-5:0.5cm) arc(-5:90 :0.5cm);     
	\fill[white!10!gray](-5 :0.5cm)--(-5:1cm) arc(-5:-105 :1cm)--(-105:0.5cm) arc(-105:-5:0.5cm);        
	\fill[orange!80!gray](-105:0.5cm)--(-105:1cm) arc(-105:-225 :1cm)--(-225:0.5cm) arc(-225:-105:0.5cm);
	\fill[blue!70!white](135:0.5cm)--(135 :1cm) arc(135:90:1cm)--(90:0.5cm) arc(90:135:0.5cm);
	\end{scope}
	\draw[thin,black](0,0) circle (0.5cm);   
	
	\shadedraw[bottom color=orange!20!gray,top color=orange!5!black,draw=black,very thin]   
	(-180:1cm) --++(0,-3mm) arc (-180:-105 :1cm) -- ++(0,3mm)  arc (-105 :-180  :1cm) -- cycle;  
	\shadedraw[bottom color=white!20!gray,top color=white!5!black,draw=black,very thin]   
	(-105:1cm) --++(0,-3mm) arc (-105:0 :1cm) -- ++(0,3mm)  arc (0 :-105  :1cm) -- cycle;  
	
	\draw[very thin] (90:0.5cm) -- (90:1cm)
	(-5:0.5cm) -- (-5:1cm)
	(-105:0.5cm) -- (-105:1cm)
	(135:0.5cm) -- (135 :1cm)
	(0,0) circle (1cm)
	(90:0.5cm)  arc (90 :135:0.5cm);
	
	\coordinate (left border) at (1.5cm,0cm); 
	\coordinate (right border) at (-1.5cm,0cm); 
	\coordinate (l1) at (43.5:0.75 cm);
	\coordinate (l2) at (-55:0.75 cm); 
	\coordinate (l3) at (117.5:0.75 cm);
	\coordinate (l4) at (-160:0.75 cm);
	
	\begin{scope}[lab/.style={gray!50!black,thick,draw}]
	\fill[lab] (l1) circle(.4mm) -- (l1-| left border) node[anchor=south east] {Till july 2020}
	node[anchor=north east] {23.52\%};         
	\fill[lab] (l2) circle(.4mm) -- (l2-| left border)  node[anchor=south east] {2019}
	node[anchor=north east] {33.83\%}; 
	\fill[lab] (l3) circle(.4mm) -- (l3-| right border) node[anchor=south west] {2018}
	node[anchor=north west] {22.06\%};         
	\fill[lab] (l4) circle(.4mm) -- (l4-| right border) node[anchor=south west] {2017}
	node[anchor=north west] {20.59\%}; 
	\end{scope}
	\end{scope}  
	\end{tikzpicture}
	\caption{Year wise article published on CSA}
	\label{fig:year}
\end{figure}

The final selection contains 68 articles (refer to Fig. \ref{Fig:a}). Reviews on the topic of hospitality and tourism were available in or after 2017 with the pace of research picking up from 2019-20, shows the increased interest of scholars in implementing ML techniques to solve CSA using online reviews in the field of hospitality and tourism challenge (refer Fig. \ref{fig:65}). Further, it can be referred from Fig. \ref{fig:year}, number of articles is increasing yearly; till July 2020, approximately published articles are the same as the completed the year of 2019.  It is also found that high impact peer-reviewed journals like Tourism Management, Decision Support Systems, Journal of Hospitality and Tourism Management, Expert systems with applications, have published many articles in this area (refer to Appendex \ref{tab:long3} for journal quality  and Table \ref{tab:long11} for number of articles). Many researchers applying ML techniques in CSA is presented in Table \ref{ml}. It is concluded that Regression, SVM, and NB,  applied by many researchers in their result evaluations. Very few researchers applied hybrid ML techniques in their work (refer to Table \ref{hybrid}). Fig. \ref{fig:11}, shows the use of the applications of ML in CSA from online reviews. The consumer sentiment analysis has 39 papers, predictive recommendation decisions having 15 papers, and fake reviews detection have 14 papers. The findings show that the ML techniques applied in all the sub-areas of CSA using online reviews in hospitality and tourism. 

\begin{landscape}
	\begin{center}
		\begin{longtable}{llll}
			\caption{Article distribution in the regards to the respective journal} \label{tab:long11} \\
			\hline \multicolumn{1}{c}{\textbf{S.No.}} & \multicolumn{1}{c}{\textbf{Name of Journal}} & \multicolumn{1}{c}{\textbf{No. of Papers}}&
			\multicolumn{1}{c}{\textbf{Proportion (\%)}}\\ \hline
			\endfirsthead
			\multicolumn{4}{c}%
			{{\bfseries \tablename\ \thetable{} -- continued from previous page}} \\
			\hline \multicolumn{1}{c}{\textbf{S.No.}} & \multicolumn{1}{c}{\textbf{Name of Journal}} & \multicolumn{1}{c}{\textbf{No. of Papers}} &
			\multicolumn{1}{c}{\textbf{Proportion (\%)}}\\ \hline 
			\endhead
			\hline \multicolumn{4}{r}{{Continued on next page}} \\ \hline
			\endfoot
			\hline \hline
			\endlastfoot
			\noalign{\smallskip}\hline\noalign{\smallskip}
			1& Tourism Management &12 &17.65\\
			2&International Journal of Hospitality Management&7&10.29\\
			3&Journal of Air Transport Management&5&7.35\\
			4&Journal of Travel Research&4&5.88\\
			5&Tourism Management Perspectives&3&4.41\\
			6&International Journal of Information Management&3&4.41\\
			7&Neurocomputing&3&4.41\\
			8&  Expert Systems& 2 &2.94\\
			9&Expert Systems With Applications&2&2.94\\
			10&Decision Support Systems&2&2.94\\
			11&Journal of Hospitality and Tourism Management&2&2.94\\
			12&Journal of Retailing and Consumer Services&2&2.94\\
			13&Journal of Business Research&2&2.94\\
			14&Information Processing and Management&2&2.94\\
			15&Information \& Management&2&2.94\\
			16& Multimedia Tools and Applications&1&1.47\\
			17&Sustainability&1&1.47\\
			18&Knowledge-Based Systems&1&1.47\\
			19&Journal of Computational Science&1&1.47\\
			20&Annals of Tourism Research&1&1.47\\
			21&Future Generation Computer Systems&1&1.47\\
			22&Telematics and Informatics&1&1.47\\
			23&Current Issues in Tourism&1&1.47\\
			24&Journal of Supercomputing&1&1.47\\
			25&Journal of Forecasting&1&1.47\\
			26&International Journal of Contemporary Hospitality Management&1&1.47\\
			27&International Journal of Intelligent Systems&1&1.47\\
			28&Information Sciences&1&1.47\\
			29&Neural Computing and Applications&1&1.47\\
			30&Applied Intelligence&1&1.47\\
			\noalign{\smallskip}\hline
		\end{longtable}
	\end{center}
\end{landscape}

\begin{table}
	\caption{ML technique covered in the SLR applied in CSA}       
	\label{ml}
	\begin{tabular}{lllp{0.45\textwidth}}
		\hline\noalign{\smallskip}
		S. No. & ML Technique & \#papers& Research references  \\
		\noalign{\smallskip}\hline\noalign{\smallskip}
		1& Clustering & 3&\cite{000000}, \cite{61}, \cite{66}\\
		2&SVM  &14 &\cite{00000}, \cite{000}, \cite{00}, \cite{0}, \cite{1}, \cite{8}, \cite{12}, \cite{14}, \cite{23}, \cite{25}, \cite{53}, \cite{54}, \cite{57}, \cite{64}\\
		3&DT  &5 &\cite{00000}, \cite{000}, \cite{00}, \cite{11}, \cite{40}\\
		4&LR  & 11&\cite{00000}, \cite{000}, \cite{2}, \cite{4}, \cite{5}, \cite{11}, \cite{14}, \cite{53}, \cite{56}, \cite{63}, \cite{64}\\
		5&LSTM&3&\cite{0000}, \cite{20}, \cite{22}\\
		6&BiLSTM&&\cite{0000}, \cite{22}\\
		7&KNN&5& \cite{12}, \cite{14}, \cite{24}, \cite{25}, \cite{57}\\
		8&NB&11&\cite{000}, \cite{0}, \cite{1}, \cite{9}, \cite{11}, \cite{12}, \cite{24}, \cite{25}, \cite{57}, \cite{66}, \cite{68}\\
		9 &Regression&16&\cite{10}, \cite{13}, \cite{15}, \cite{19}, \cite{37}, \cite{41}, \cite{43}, \cite{44}, \cite{47}, \cite{48}, \cite{49}, \cite{52}, \cite{58}, \cite{59}, \cite{62}, \cite{67}\\
		10&CNN&5&\cite{18}, \cite{20}, \cite{22}, \cite{25}, \cite{55}\\
		11&RF&6&\cite{000}, \cite{00}, \cite{8}, \cite{11}, \cite{14}, \cite{23}\\
		12&LDA&5&\cite{22}, \cite{33}, \cite{39}, \cite{42}, \cite{67}\\
		13&GB&2&\cite{000}, \cite{14}\\
		14&NN&8&\cite{000}, \cite{00}, \cite{1}, \cite{7}, \cite{8}, \cite{18}, \cite{20}, \cite{71}\\ 
		\noalign{\smallskip}\hline
	\end{tabular}
\end{table}

\begin{table}
	\caption{Year Wise distribution of applied hybrid technique covered in the literature}       
	\label{hybrid}
	\begin{tabular}{llll}
		\hline\noalign{\smallskip}
		S.No. & Hybrid Technique &Year& Research references  \\
		\noalign{\smallskip}\hline\noalign{\smallskip}
		1&RBFNN  &2020  &\cite{0000}\\
		2& ANN+ Firefly, ANN+ Bat &2018  &\cite{7}\\
		3&  PCA+LGR PCA+SVM &2018&\cite{40}\\
		4&PCA+Biplot Regression&2017&\cite{43}\\
		\noalign{\smallskip}\hline
	\end{tabular}
\end{table}
\definecolor{strangegreen}{RGB}{15,111,83}
\pgfkeys{%
	/piechartthreed/.cd,
	scale/.code                =  {\def\piechartthreedscale{#1}},
	mix color/.code            =  {\def\piechartthreedmixcolor{#1}},
	background color/.code     =  {\def\piechartthreedbackcolor{#1}},
	name/.code                 =  {\def\piechartthreedname{#1}}}

\newcommand\piechartthreed[2][]{%
	\pgfkeys{/piechartthreed/.cd,
		scale            = 1,
		mix color        = gray,
		background color = white,
		name             = pc} 
	\pgfqkeys{/piechartthreed}{#1}
	\begin{scope}[scale=\piechartthreedscale] 
		\begin{scope}[xscale=5,yscale=3] 
			\path[preaction={fill=black,opacity=.8,
				path fading=circle with fuzzy edge 20 percent,
				transform canvas={yshift=-15mm*\piechartthreedscale}}] (0,0) circle (1cm);
			\pgfmathsetmacro\totan{0} 
			\global\let\totan\totan 
			\pgfmathsetmacro\bottoman{180} \global\let\bottoman\bottoman 
			\pgfmathsetmacro\toptoman{0}   \global\let\toptoman\toptoman 
			\begin{scope}[draw=black,thin]
				\foreach \an/\col [count=\xi] in {#2}{%
					\def\space{ } 
					\coordinate (\piechartthreedname\space\xi) at (\totan+\an/2:0.75cm); 
					\ifdim 180pt>\totan pt 
					\ifdim 0pt=\toptoman pt
					\pgfmathsetmacro\toptoman{180} 
					\global\let\toptoman\toptoman 
					\else
					\fi
					\fi   
					\fill[\col!80!gray,draw=black] (0,0)--(\totan:1cm)  arc(\totan:\totan+\an:1cm)
					--cycle;     
					\pgfmathsetmacro\finan{\totan+\an}
					\ifdim 180pt<\finan pt 
					\ifdim 180pt=\bottoman pt
					\shadedraw[left color=\col!20!\piechartthreedmixcolor,
					right color=\col!5!\piechartthreedmixcolor,
					draw=black,very thin] (180:1cm) -- ++(0,-3mm) arc (180:\totan+\an:1cm) 
					-- ++(0,3mm)  arc (\totan+\an:180:1cm);
					\pgfmathsetmacro\bottoman{0}
					\global\let\bottoman\bottoman
					\else
					\shadedraw[left color=\col!20!\piechartthreedmixcolor,
					right color=\col!5!\piechartthreedmixcolor,
					draw=black,very thin](\totan:1cm)-- ++(0,-3mm) arc(\totan:\totan+\an:1cm)
					-- ++(0,3mm)  arc(\totan+\an:\totan:1cm); 
					\fi
					\fi
					\pgfmathsetmacro\totan{\totan+\an}  \global\let\totan\totan 
				} 
			\end{scope}
		\end{scope}  
	\end{scope}
}

\section{Review findings and discussions}
In this section, we evaluate and present the conclusion of SLR in three sub-area. These sub-area represents the use of ML in CSA applications such as sentiment analysis \cite{136}, predictive recommendation \cite{137}, and fake reviews detection \cite{138}. The use of ML in each sub-area serves a particular application for the CSA process. It is found from the SLR that ML is implemented in the consumer sentiment analysis in the field of hospitality and tourism management using online reviews (sub-area I). 

Similarly, ML techniques are used in predictive consumer recommendation decisions based on the previous consumer experience regarding offered services (sub-area II).  In this sub-area II, the use of ML is for overall consumer satisfaction and their recommendation for the forthcoming consumers. Fake review detection (sub-area III) is the major application of ML in identify reviews that are misguiding new consumer before making their purchase decision. 

\subsection{Sentiment analysis}
Sentiment analysis or opinion mining is the process of understanding and classifying consumer feelings or emotions (positive, negative, or neutral) within reviews using text analysis, ML, and NLP \cite{139}. Sentiment analysis models allow organizations to identify consumer satisfaction towards their products, services, or offerings in online reviews. CSA is essential for forthcoming consumers and business organizations in making their decisions efficiently.   

With the advancement of ICT, user-generated content on the online platform has made sentiment analysis an essential tool for disentangling consumer sentiment information about a product or service. Recent research focuses on sentiment analysis for two separate objectives; first is aspect extraction, and sentiment classification of offerings, and another is sentiment classification of targeted tweets. ML techniques have emerged as a prospect for obtaining such goals with their capability to capture the textual features without requirements of high-level feature extraction. Sherwin Shenwei \&Lim, Aaron Tkaczynski \cite{41} used regression in explaining the importance of origin and money in consumer expectations. Lee et al. \cite{40} use DT, RF, PCA in the identification of online reviews helpfulness in consumer satisfaction. Similarly, many researchers implemented ML techniques over various data in finding their objectives. Different data set used in sentiment analysis shown in Table \ref{sentiment}. Table \ref{sentiment} presents that most of the researchers are using hotel, airline, and tourism reviews in their study, and there is a future scope in airport and museum reviews analysis.

\begin{table}
	\caption{Widely used data sets for Sentiment Analysis}   
	\label{sentiment}    
	\begin{tabular}{lllp{0.45\textwidth}}
		\hline\noalign{\smallskip}
		S. No. & Dataset used & \#papers&Research references  \\
		\noalign{\smallskip}\hline\noalign{\smallskip}
		1& Hotel reviews & 13&\cite{31},\cite{33},\cite{34},\cite{40},\cite{53},\cite{55}, \cite{56},\cite{57},\cite{58},\cite{61},\cite{62},\cite{63}, \cite{64} \\
		2& Airline reviews &12& \cite{36},\cite{37},\cite{38},\cite{39},\cite{41},\cite{42}, \cite{43},\cite{44},\cite{45},\cite{48},\cite{49},\cite{50} \\
		3& Restaurant Reviews&5 &\cite{52}, \cite{54},  \cite{57}, \cite{59}, \cite{65} \\
		4& Airport Reviews & 2&\cite{47},\cite{60} \\
		5& Tourist Rewiews & 1& \cite{51}\\
		6& Art and museum Reviews&1& \cite{57} \\
		\noalign{\smallskip}\hline
	\end{tabular}
\end{table}
\subsection{Predictive consumer recommendation decision}
For any organization, consumer reviews are important, not only for identifying factors that are responsible for business growth and performance, but also for the improvement of offered services or products as well as for enhancement of consumer feeling or opinion. In most of the cases, organizational management trust on consumer reviews factors such as overall consumer satisfaction, percentages of consumer complaints, repurchasing of items as well as word-of-mouth like recommendations or promotions \cite{140}. 

In predictive consumer recommendations, Reichheld \cite{141} suggested the convention of the word of mouth recommendations for offerings, associated with the net promoter score (NPS). He advised that NPS is the most reliable parameter in the prediction of business growth in comparison with the other measurements of consumer satisfaction.  NPS is calculated as the percentage difference of the promoters and detractors. In the past few years, many researchers included NPS as a consumer satisfaction measure in their study. They found that business growth is directly proportional to NPS \cite{141}. From the Reichheld \cite{141} finding, it can be concluded that NPS is the best measure of consumer satisfaction and predictive consumer recommendation decision \cite{142}. In organizations, NPS considered as a base for consumer satisfaction and drawing parameters for business growth \cite{143}. Hence, processes based on the NPS are particularly efficient in sensing and understanding the overall method of gaining to consumer and substantiating choice \cite{143}. 

Given the significance of consumer involvement, consumer loyalty, and consumer reviews, as reflected in NPS, this is also important to understand the parameters that are responsible for positive word-of-mouth. Nowadays, consumers are expressing their feeling or opinion regarding their experience in online reviews; organizations need to understand the consumer preferences and factors influencing their predictive suggestions. Nonetheless, many organizations are not capable of using consumer reviews to convert it into business growth \cite{144}. This is due to that some of the reviews are not providing consumer recommendations directly, i.e., the issue of whether a consumer is suggesting a particular service for the forthcoming consumers or not \cite{145}. Thus, this is a research gap in automatically scrapping information from user-generated online reviews to accurately predicting drivers for consumer recommendations, understanding the advice to different services, and finally forming them into NPS. 

In current research work, many authors predicting consumer recommendations decision based on online reviews, Siering et al. \cite{146} used qualitative content form online reviews in explaining and predicting airline consumer recommendation decisions, whereas Chatterjee et al. \cite{147} utilized qualitative and quantitative content in predicting consumer recommendation decisions. Many researchers applied machine learning techniques in predicting consumer recommendations in various fields (refer to Table \ref{rec}). Form Table \ref{ml}, we may refer to that many scholars used regression and basic machine learning techniques in their study. For the future work, ensembling and optimization procedures may apply for getting predictive recommendation decisions.

\begin{table}[!h]
	\caption{Widely used data sets for Predictive recommendation decision}       
	\label{rec}
	\begin{tabular}{llll}
		\hline\noalign{\smallskip}
		Rank & Dataset used & \#papers&Research references  \\
		\noalign{\smallskip}\hline\noalign{\smallskip}
		1&Hotel  Review  &6 &\cite{000000}, \cite{00000}, \cite{000}, \cite{00}, \cite{8}, \cite{9} \\
		2&  Tourism Reviews& 1&\cite{0000}\\
		3&Airline reviews&5&\cite{1}, \cite{2}, \cite{3}, \cite{5}, \cite{6}\\
		4&Restaurant reviews&1&\cite{4}\\
		5&Trip reviews&1&\cite{7}\\
		\noalign{\smallskip}\hline
	\end{tabular}
\end{table}
\subsection{Fake reviews detection}

A fake review is a positive or negative review provided for a product or service by anyone who may not even experience the service or did not purchased any product, but (s)he is writing reviews. A fake reviewer may be a seller or any person on behalf of the seller, providing reviews using fake credentials, they are posting a fake image of the product and encouraging consumers to buy such products by advertising falsely. Fake reviewers are providing fake reviews for their own benefit by providing negative reviews for their competitators and positive reviews for their own products.
 
There might be many reasons providing fake reviews for products or services, but in short, the main goal is to sell more products or degrade the selling of products of competitors. Some reasons are mentioned here. 

\begin{enumerate}
	\item To improve the product’s ranking
	\item	To improve the seller’s ranking
	\item	To boost sales of low-selling products
	\item	To boost the visibility of products which have just been listed
	\item	To balance negative reviews left for a product
	
\end{enumerate}
Online consumer reviews regarding products or services are playing a vital role for consumers, creating a different type of Word-of-mouth information \cite{148}. Form the current research, and it is found that 52\% of online consumers are using online reviews to find product pieces of information, while 24\% of them are using previous consumer reviews before making their purchase decision \cite{149}. Consequently, online reviews have strong influences on consumers' decision making in purchasing a product or services, there are many related areas affected by it, some of them are hospitality and tourism \cite{150,151}, online purchase \cite{152}, and entertainment \cite{153,154,155}. Moreover, many platforms are providing online reviews for the same service or product, which may be classified \cite{156} as per the platform that is giving the word of mouth information. 

 Articles based on the fake reviews detection included in this SLR process shown in Table \ref{fake}. 
 
 \begin{table}[!h]
 	\caption{Widely used data sets for Fake reviews detection}       
 	\label{fake}
 	\begin{tabular}{lllp{0.45\textwidth}}
 		\hline\noalign{\smallskip}
 		S.No. & Dataset used & \#papers&Research references  \\
 		\noalign{\smallskip}\hline\noalign{\smallskip}
 		1& Restaurant Reviews & 11&\cite{10}, \cite{12}, \cite{13}, \cite{15}, \cite{17},  \cite{18}, \cite{19}, \cite{20}, \cite{22}, \cite{23}, \cite{25} \\
 		2&Hotel Reviews  &5&\cite{12}, \cite{14}, \cite{18}, \cite{20}, \cite{22}, \cite{25} \\
 		3&Monuments reviews&1&\cite{21}\\
 		\noalign{\smallskip}\hline
 	\end{tabular}
 \end{table}
\section{Proposed framework and implications}
\subsection{ML-CSA framework}
The SLR findings show that ML has a vast possibility for application in CSA using online reviews in hospitality and tourism. The data gathered from the various online platform in CSA is utilized for making classification and predictive recommendation decisions using ML techniques. The findings show that ML techniques improve the accuracy of CSA and address the various challenges faced during sentiment analysis, fake reviews detection, and predictive consumer recommendation decisions. The use of ML techniques also helps in drawing valuable aspects responsible for consumer sentiment. We utilized findings from the SLR to design an ML-CSA framework that may be helpful for researches in drawing their objective. 

\begin{landscape}
	\begin{figure}
		\centering
		\includegraphics[width=18cm, height=12cm]{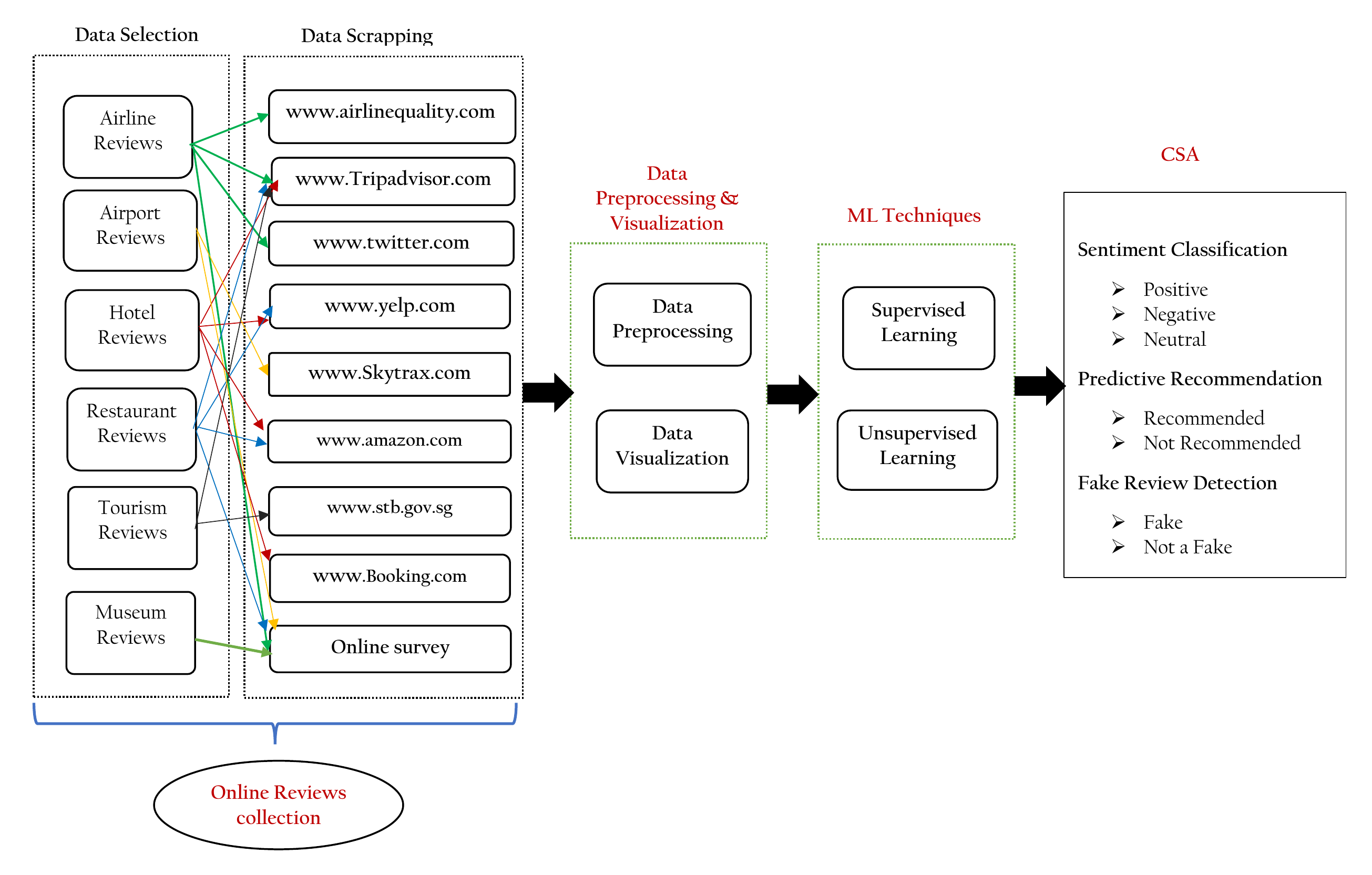}
		\caption{A ML-CSA framework}
		\label{ML-CSA}       
	\end{figure}
\end{landscape}

We proposed framework in Fig \ref{ML-CSA}. has four phases, online reviews collection, data preprocessing and visualizations, ML techniques, and CSA. 

\subsubsection{Online reviews collection}
Nowadays, consumers are increasingly relying on online reviews as an essential source of information before making their purchase decision \cite{157}. In this case, many researchers found the importance of online reviews can be seen as a clue of information through the various phases of the purchase decision-making process \cite{158} and that demand for the services that are reviewed by the previous consumers \cite{159}. Further, online reviews are more significant for online retailers for attracting consumers, and next time they might purchase the reviewed item. 

There are various domains of research focused on word of mouth within the online reviews. Existing literature represents a variety of research relating word of mouth to the sales. In contrast, this research finds a significant impact of online reviews on demand for particular items \cite{160,161}. In another field, researchers focused on the factors that are making online reviews effective and creditable. In this case, reviews writing style \cite{165}, reviews contents \cite{162,163,164}, and community membership \cite{166} have been found most relevant. From an experimental point of view, drawing of the attributes impacting online reviews creditable may be used to improve consumer policies \cite{167}. In another stream of research, researchers consentrated on the reviewers contribution and factors that are motivating consumers to write their reviews on the various available platform \cite{168,169}. 

Research mainly focusing on hospitality and tourism show that forthcoming consumer can consider online reviews as a source of information before making purchase decisions. Current surveys amoung touristers advises that a total of 20-45\% of tourists are using online reviews provided by previous consumer to suggest or guide while making their decision-making process \cite{170,171}. An author \cite{172} in,  have listed the behaviour of hotel guest feeling mentioned in online reviews and explaining the role of hidden features of hotel guest sentiment in understanding guest satisfaction. Many of authors find the importance of online reviews in explaining and predicting consumer recommendation nature.

\begin{figure}
	\centering
	\includegraphics[width=12cm, height=7cm]{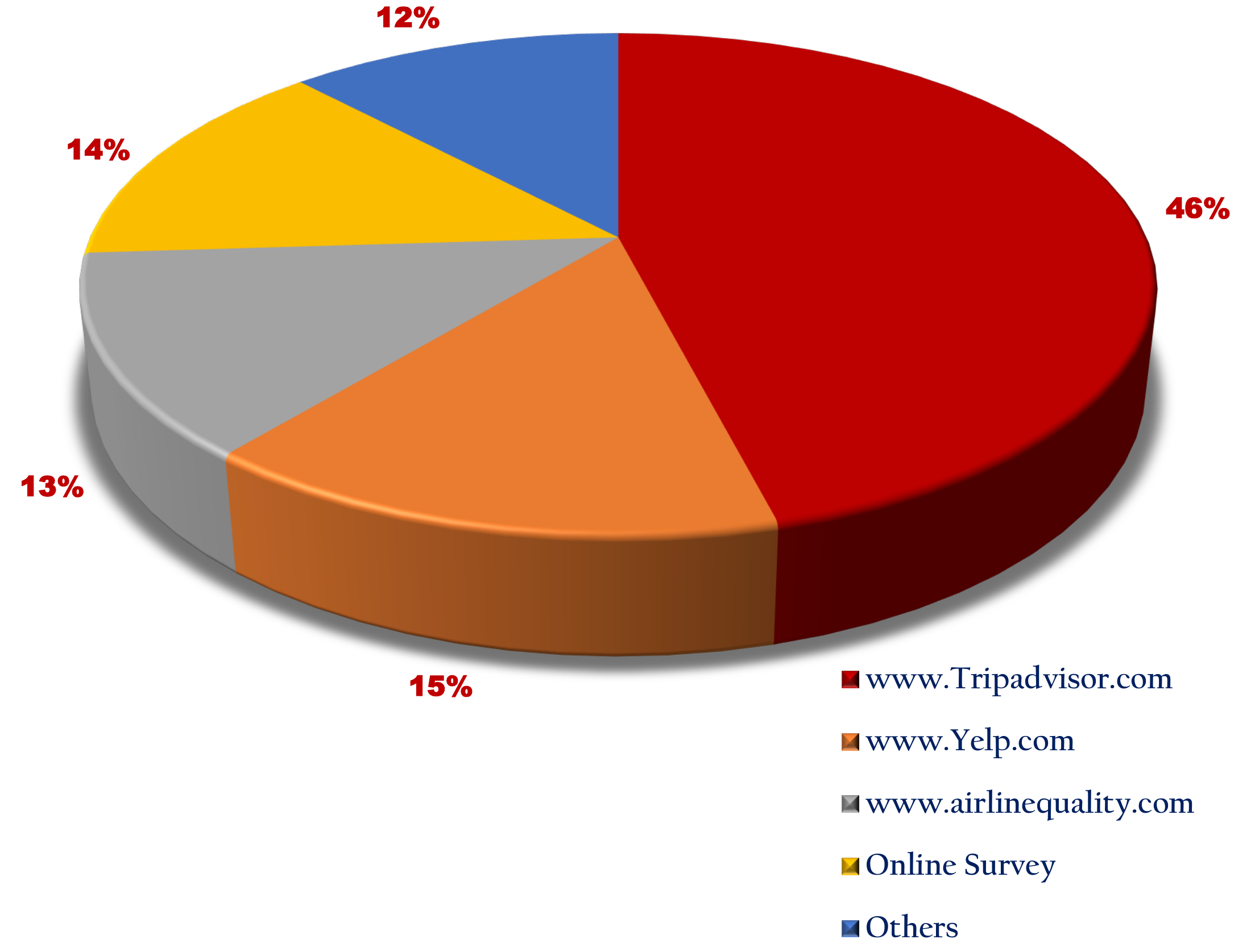}
	\caption{Please write your figure caption here}
	\label{per}       
\end{figure}

The first phase in this framework is data selection as per the research goal such as airline reviews, airport reviews, hotel reviews, etc. A sufficient amount of data gets collected by applying a data scraping procedure. Tripadvisor, Skytrax, and yelp, etc. these are the various websites that are providing online reviews data related to tourism and hospitality.  We have mentioned different dataset used in the existing literature in Table \ref{Data-source} with the data source. We have also presented the percentage of the data source contribution to this SLR related articles (refer to Fig. \ref{per}). 

\begin{table}[!h]
	\caption{Widely used data sets and sources used for CSA} 
	\label{Data-source}      
	\begin{tabular}{llp{0.35\textwidth}}
		\hline\noalign{\smallskip}
		S. No. & Dataset used & Data source  \\
		\noalign{\smallskip}\hline\noalign{\smallskip}
		1& Hotel reviews &\url{www.booking.com}, \url{www.Tripadvisor.com}, \url{www.yelp.com}, \url{www.amazon.com}\\
		2& Airline reviews & \url{www.airlinequality.com}, \url{www.Tripadvisor.com}, 
		Online survey\\
		3& Restaurant Reviews &\url{www.Yelp.com}, \url{www.Tripadvisor.com}, Online survey   \\
		4& Airport Reviews & \url{www.skytrax.com}, \url{www.twitter.com} 
		
		Online survey\\
		5& Tourist Rewiews & \url{www.Tripadvisor.com}, \url{https://www.stb.gov.sg}\\
		6& Art and museum Reviews& London Art Museums reviews\\
		\noalign{\smallskip}\hline
	\end{tabular}
\end{table}  

\subsubsection{Data Preprocessing and visualizations}
Online reviews data collected from the websites are mostly in unstructured form, but the ML technique works efficiently with the structured data; therefore, there is a need to convert unstructured data into structured data. Processing of converting unstructured data into structured one with the help of Natural language processing and text mining approaches referred to as data preprocessing. This text mining process helps in extracting consumer sentiment form online reviews. There are many areas where text mining has been applied; some of them are topic modelling, sentiment or opinion analysis, text summarization etc. 

Text preprocessing is an essential phase of data mining, in which data convert into a usable form for applying ML techniques. Text preprocessing task with their descriptions presented in Table \ref{tab:33}.

\vspace{-.4cm}

\subsubsection{ML techniques}
The third phase in the proposed framework is the selection of various ML techniques that may be applied in CSA. The SLR has found applicability of supervised and unsupervised learning techniques is utilized with the focus of sentiment analysis, predictive recommendation, and fake reviews detection. There is a possibility of the applicability of reinforcement learning in CSA. Our recommended framework advises that the CSA and ML techniques have a bridge that mainly focused on utilizing the power of ML techniques to drawing the factors influencing consumer sentiment by applying appropriate ML techniques. Further it is also suggested that ML techniques have already been applied and in future work which ML techniques can be implemented in CSA.  
\subsubsection{Decision making}
The SLR findings report that the objective of analyzing online reviews data by applying ML techniques is to draw the consumer sentiment in the field of hospitality and tourism. It was amusing to illustrate that the CSA utilizes ML techniques not only to classify consumer sentiment but also for predictive consumer recommendation decisions and fake reviews detection. Section 4 has presented the importance of ML techniques in enabling CSA in the field of hospitality and tourism. In the case of the purchase decision, the power of ML techniques in forecasting the previous consumer recommendations about products and services \cite{7}. The hybrid ML models can help in predicting consumer sentiment accurately \cite{13,41}.  Identifications of the theme shared in the user-generated online reviews by traveler and linking of that theme with the values of money are presented in \cite{36}. The influences of the various aspects of airline service failures and the respective recovery action on consumer sentiment also studied in literature \cite{37}. 

From the above discussion, we may conclude that the decision making the last step of the CSA process, so we have included it in our proposed ML-CSA framework. The various performance measures are described in Table \ref{KPI}, this table also shows the uses of performance measures applied in the articles selected for SLR.  


	\begin{center}
		\begin{longtable}{p{0.08\textwidth}p{0.15\textwidth}p{0.25\textwidth}p{0.1\textwidth}p{0.30\textwidth}}
			\caption{KPI Descriptions }
		\label{KPI}  \\
			\noalign{\smallskip}\hline\noalign{\smallskip}
			\hline\multicolumn{1}{c}{\textbf{S.No.}} & \multicolumn{1}{c}{\textbf{KPI}} & \multicolumn{1}{c}{\textbf{Equation}}&
			\multicolumn{1}{c}{\textbf{\# Papers}}& 
			\multicolumn{1}{c}{\textbf{Research  References}}\\ \hline
			\endfirsthead
			\multicolumn{5}{c}%
			{{\bfseries \tablename\ \thetable{} -- continued from previous page}} \\
			\hline \multicolumn{1}{c}{\textbf{S.No.}} & \multicolumn{1}{c}{\textbf{KPI}} & \multicolumn{1}{c}{\textbf{Equation}}&
		\multicolumn{1}{c}{\textbf{\# Papers}}& 
		\multicolumn{1}{c}{\textbf{Research  References}}\\ \hline 
			\endhead
			\hline \multicolumn{5}{r}{{Continued on next page}} \\
			\hline
			\endfoot
			\hline \hline
			\endlastfoot
			\noalign{\smallskip}\hline\noalign{\smallskip}
		1&RMSE&$\sqrt{\frac{1}{N}\sum_{t=1}^{N}\left( x_{t}-\hat{x_{t}}\right)^{2}}$& 6& \cite{000000}, \cite{0000}, \cite{000}, \cite{0}, \cite{7}, \cite{63}   \\
		2&MAE&$\frac{1}{N}\sum_{t=1}^{N}\left|x_{t}-\hat{x_{t}}\right|$&6&\cite{0000}, \cite{000}, \cite{7}, \cite{8}, \cite{63}, \cite{71}   \\
		3&MAPE&$\frac{1}{N}\sum_{t=1}^{N}\left| \frac{x_{t}-\hat{x_{t}}}{x_{t}} \right|$&4& \cite{0000}, \cite{000}, \cite{7}, \cite{8}   \\
		4&$R^2$&$1-\frac{1}{N}\frac{\sum_{t=1}^{N}(x_{t}-\hat{x_{t}})^2}{\sum_{t=1}^{N}(x_{t}-\bar{x_{t}})}$&3
		&  \cite{000000}, \cite{4}, \cite{52}   \\
		5&MdAPE&$Median\left| \frac{x_{t}-\hat{x_{t}}}{x_{t}} \right|$&2&\cite{8}, \cite{71}\\
		6&Precision&$\frac{TP}{TP+FP}$&14&\cite{00000}, \cite{00}, \cite{0}, \cite{1}, \cite{14}, \cite{20}, \cite{22}, \cite{23}, \cite{31}, \cite{40}, \cite{54}, \cite{57}, \cite{64}, \cite{68}  \\
		7&Recall or Sensitivity&$\frac{TP}{TP+FN}$& 14& \cite{00000}, \cite{00}, \cite{0}, \cite{1}, \cite{17}, \cite{20}, \cite{22}, \cite{23}, \cite{31}, \cite{40}, \cite{54}, \cite{55}, \cite{64}, \cite{64}
		 \\
		8&F-Measure&$2*(\frac{Precision*Recall}{Precision+Recal})$&16&  \cite{00000}, \cite{00}, \cite{0}, \cite{1}, \cite{12}, \cite{14}, \cite{17}, \cite{18}, \cite{20}, \cite{22}, \cite{23}, \cite{25}, \cite{40}, \cite{54}, \cite{55}, \cite{68}  \\
		9&Accuracy&$\frac{TP+FP}{TP+FP+TN+FN}$& 15& \cite{000}, \cite{00}, \cite{0}, \cite{1},  \cite{10}, \cite{17}, \cite{18}, \cite{20}, \cite{22} \cite{25},  \cite{40}, \cite{46} \cite{54}, \cite{57}, \cite{68} \\
		10&AUC&&3&  \cite{00}, \cite{23}, \cite{24}  \\
		11&Specificity&$\frac{TN}{TN+FP}$& 2& \cite{00}, \cite{40}  \\
			\noalign{\smallskip}\hline
\end{longtable}
\end{center}

\subsection{Implications of SLR}
\subsubsection{Implications for researchers}
\label{5.2.1}
We tryied to answer all queries of researchers who are new or already working in the area of application of ML in tourism and hospitality (refer to Table \ref{fut}).   Based on the analysis of this survey, we also present the mentioned future scope of CSA using ML in hospitality and tourism.  
\begin{enumerate}
	\item The SLR focuses on ML applications in the analysis of online reviews written in English, and no attention is paid to written reviews in another language; further work is required in the analysis of written reviews in different languages, such as Hindi, Parsi, Urdu, Bengali, etc.
	\item The findings show that ML techniques help to improve the performance of the model applied in CSA. However, it would be impressive to draw the factors affecting the performance of the model. Future research should focus on measuring the influence of CSA and provide particular guidance on how CSA can be evaluated by applying ML techniques. The ML-CSA proposed framework in this SLR can be implemented as a guideline framework for this type of research work.  
	\item Future research should focus on first detecting fake reviews and then applying CSA to the remaining data. 
	\item CSA Software and tools are not available publicly; in the future, researchers can develop such intelligent and personalize CSA tools and softwares, by which users can analyze previous consumer's sentiment before making their purchase decision. 
	\item Though many researchers are efficient in applying ML techniques, only few of them used technologies such as optimization algorithms, ensemble learning, deep learning, and neuro-fuzzy models in CSA; the researcher may include such techniques in their future work. 
	\item Analyzing CSA using Fuzzy Logic methods like Type 2 Fuzzy models is yet another novel direction that is open for further exploration.
	\item Very few researchers considered museum reviews dataset in their analysis; in the future, this can also be scope for further study.
	\item Due to the uavailability of online reviews on consumer experiences in various historical places, no article found yet on the application of ML in this area, reserchers may analysis historical online reviews data using ML techniques in finding consumer satisfaction and predictive recommendation decisions. 
	\item Most of the work included qualitative online reviews in their study, whereas qualitative and quantitative review content both added by a few researchers in their study. 
	\item Consumer choice of preferences as per their region and culture have been found in very few articles; this can also be a future direction of research. 
	\item The findings of this SLR indicate that researchers are using supervised and unsupervised ML techniques in analyzing CSA, reinforcement ML techniques can be implemented in future research.  
	\item The results of this SLR suggest that researchers are still using ML techniques to analyze consumer sentiment on online textual reviews of data only. There is potential scope for using ML techniques to analyze consumer sentiment on the online reviews data in the form of audio, image, and video, etc.  
\end{enumerate}
\begin{landscape}
\begin{table}[!h]
	\caption{Research Scholars queries and their solutions form SLR findings}
	\label{fut}       
	\begin{tabular}{p{0.08\textwidth}p{0.6\textwidth}p{0.7\textwidth}p{0.4\textwidth}}
		\hline\noalign{\smallskip}
		S. No. & Research Scholars Queries & Response from findings& Reference scetion  \\
		\noalign{\smallskip}\hline\noalign{\smallskip}
		1& What is sentiment analysis? &Drawing consumer experience from their reviews, consumer experience regarding services may be positive, negative, or neutral.& Subsection \ref{2.1}   \\
		2& What are the challenges of consumer sentiment analysis? & Few challenges are subjectivity detection, computational cost, context-dependency, and accuracy. & Subsection\ref{2.2}\\
		3& How the study of consumer sentiment analysis can be useful for consumers. &Consumer can refer to previous consumer experience before making their purchase decisions. & Section \ref{intro}\\
		4& How the study of consumer sentiment analysis can be useful for organizations. &Organizations can refer to consumer experience in the making of consumer policy and product improvement. & Section \ref{intro} \\
		5&From where consumer reviews get collected?  &There are many online platforms available for online reviews.& Table \ref{Data-source}\\
		6& How can online reviews be preprocessed? &Online reviews can be preprocessing with the help of text mining and natural language processing techniques. & Table \ref{tab:22} \&Table \ref{tab:33} \\
		
		7&What are the techniques used in CSA?  & Most of the researchers preferred ML techniques in their implementation. & Table \ref{ml}\\
		8&What are the performance measures used in CSA?  & There are many performance matrices preferred by researchers such as precision, recall, and accuracy, etc. & Table \ref{KPI}\\
		9& What is Future scope in CSA? &In future research work, the researcher can include culture variables while predicting consumer satisfaction. Some future direction is mentioned in implications for the researcher's section. & Subsection \ref{5.2.1}  \\
		
		\noalign{\smallskip}\hline
	\end{tabular}
\end{table}
\end{landscape}
\subsubsection{Implications for service provider}
This SLR gives many managerial implications, which may be useful for any organization for serving its consumers in a better way. The primary findings of this SLR are the importance of the contents of online reviews in making consumer decisions. Our findings show that core service aspects have greater influences on consumer sentiments than augmented service aspects. We have also seen that consumer sentiment or emotion can also explain predictive consumer recommendations; this is providing valuable suggestions in designing service strategy for their consumers. A literature survey also represents that the consumer sentiment is context-dependent; whenever drawing consumer sentiment, Management has to be careful about the context. While designing the advertisement content, they have to be attentive regarding sentences they are using, such as sentimental words related to joy having much impact than surprises or trust.  Consumer choices of services are dependent on various factors such as region and culture from which they belong to, cost, and facilities. Therefore, Management has to be attentive in designing service policies as per the consumers' choice.  
\section{Summery}

The current study is based on an SLR to determine the recent state studies on ML techniques in CSA. The SLR conducted on 68 research papers, which were based on sentiment classification, predictive recommendation decisions, and fake reviews detection in the field of hospitality and tourism. The SLR finds that supervised and unsupervised machine learning techniques are used in CSA based on online reviews. The novelty of this SLR is the proposed ML-CSA framework; this is providing valuable guidelines to new and experienced researchers. Our proposed framework is providing complete research information from topic selection to publications such as basic knowledge about CSA, data selection, data sources, data preprocessing, feature selection, applied ML techniques, current research objective, and future scope, etc. Under our knowledge, no SLR providing guidelines about journal selection and required revision time by the selected journal. This SLR is also providing journal descriptions related to the area of SLR objective; this will be a great help for new researchers who start working in this area.  The SLR revels considerable importance to the CSA that has achieved by the power of ML techniques concludes that the implementation of ML in CSA is beneficial. Considering the online reviews and capability of ML techniques, the service provider can draw factors consumer satisfaction and utilize findings in their business growth. Additionally, the forthcoming consumer can also refer to previous consumer experience regarding services or products before making their purchase decisions. 

Like other studies, our study also has some limitations. We tried to include all research articles in our SLR, but there is a possibility, we might missed a few important research articles. The SLR captured articles published in a timeframe from Jan 2017 to July 2020, i.e. current research articles. The list of articles included in  SLR, as they are selected from top journals in this field or indexed by SSCI/SCI/SCIE, but the chance is there for any important article published in any conference might be excluded. Therefore, future studies can include articles from top conferences and from other databases. The proposed ML-CSA framework is based on the finding from the current research, which empirically has not been tested. Hence, validation of the proposed framework can be included in future work. Additionally, future research may also explore the power of ML to CSA in different domains such as academics, business, military surveillance, etc.

\appendix

\begin{landscape}
	
	\section{\hspace{.2cm}Journals description from which articles selected for SLR}
	\begin{center}
		\begin{longtable}{lllllll}
			\caption{Descriptions of Journals included in SLR\\
				Quartile and Indexing as per JCR Web of Science 2020\\
				Min(Minimum), Max(Maximum), and Avg (Average) represents the No. of days taken by that joural for the articles included in SLR} \label{tab:long3} \\
			\hline \multicolumn{1}{c}{\textbf{S.No.}} & \multicolumn{1}{c}{\textbf{Name of Journal}} & \multicolumn{1}{c}{\textbf{Quartile}}&
			\multicolumn{1}{c}{\textbf{Indexing}}& 
			\multicolumn{1}{c}{\textbf{Min}}& 
			\multicolumn{1}{c}{\textbf{Avg}}&
			\multicolumn{1}{c}{\textbf{Max}}\\ \hline
			\endfirsthead
			\multicolumn{7}{c}%
			{{\bfseries \tablename\ \thetable{} -- continued from previous page}} \\
			\hline \multicolumn{1}{c}{\textbf{S.No.}} & \multicolumn{1}{c}{\textbf{Name of Journal}} & \multicolumn{1}{c}{\textbf{Quartile}} &
			\multicolumn{1}{c}{\textbf{Indexing}}& 
			\multicolumn{1}{c}{\textbf{Min}}& 
			\multicolumn{1}{c}{\textbf{Avg.}}&
			\multicolumn{1}{c}{\textbf{Max}}\\ \hline 
			\endhead
			\hline \multicolumn{7}{r}{{Continued on next page}} \\
			\hline
			\endfoot
			\hline \hline
			\endlastfoot
			\noalign{\smallskip}\hline\noalign{\smallskip}
			1& Tourism Management &Q1 &SSCI&2&233.67&619\\
			2&International Journal of Hospitality Management&Q1&SSCI&110&244&353\\
			3&Journal of Air Transport Management&Q2&SSCI&132&180&218\\
			4&Journal of Travel Research&Q1&SSCI&-&-&-\\
			5&Tourism Management Perspectives&Q1&SSCI&45&122&219\\
			6&International Journal of Information Management&Q1&SSCI&117&163&190\\
			7&Neurocomputing&Q1&SCIE&180&334&428\\
			8&  Expert Systems& Q2 &SCIE&152&320&488\\
			9&Expert Systems With Applications&Q1&SCIE&81&128&174\\
			10&Decision Support Systems&Q1&SCIE&108&161&214\\
			11&Journal of Hospitality and Tourism Management&Q2&SSCI&222&238&253\\
			12&Journal of Retailing and Consumer Services&Q2&SSCI&86&202&318\\
			13&Journal of Business Research&Q1&SSCI&273&311&348\\
			14&Information Processing and Management&Q1&SCIE&178&252&325\\
			15&Information \& Management&Q1&SCIE&211&314&416\\
			16& Multimedia Tools and Applications&Q2&SCIE&-&236&-\\
			17&Sustainability&Q2&SCIE&-&50&-\\
			18&Knowledge-Based Systems&Q1&SCIE&-&270&-\\
			19&Journal of Computational Science&Q2&SCIE&-&166&-\\
			20&Annals of Tourism Research&Q1&SSCI&-&135&-\\
			21&Future Generation Computer Systems&Q1&SCIE&-&382&-\\
			22&Telematics and Informatics&Q1&SSCI&-&124&-\\
			23&Current Issues in Tourism&Q1&SSCI&-&260&-\\
			24&Journal of Supercomputing&Q2&SCIE&-&-&-\\
			25&Journal of Forecasting&Q2&SSCI&-&251&-\\
			26&International Journal of Contemporary Hospitality Management&Q1&SSCI&-&364&-\\
			27&International Journal of Intelligent Systems&Q1&SCIE&-&-&-\\
			28&Information Sciences&Q1&SCIE&-&299&-\\
			29&Neural Computing and Applications&Q1&SCIE&-&75&-\\
			30&Applied Intelligence&Q2&SCIE&-&-&-\\
			\noalign{\smallskip}\hline
		\end{longtable}
	\end{center}
\end{landscape}

\noindent\textbf{Author contribution}\\
The author approves their contribution in this study as follows, study concept and design, data collection, literature review, proposed model; Praphula Kumar Jain,   drafting the manuscript; Praphula Kumar Jain, edit and English corrections Rajendra Pamula. All authors went through the final version of the paper and approved this manuscript.\\ 

\noindent\textbf{Acknowledgments}\\
This research work has completed at the Department of Computer Science \& Engineering, Indian Institute of Technology (Indian School of Mines), Dhanbad, JH, INDIA. We are thankful for the institutional support and providing a computer lab facility.  We will also be thankful for esteemed reviewers and editors for their helpful suggestions in the improvement of this manuscript and getting published in a reputed journal. \\

\noindent\textbf{Compliance with Ethical Standards}\\

\noindent\textbf{Conflict of interest} All authors declare that they have no conflict of interest.\\

\noindent\textbf{Ethical approval} This article does not contain any studies with human participants or animals performed by any of the authors.\\

\noindent\textbf{Funding Information} This research received no specific grant from any funding agency in the public, commercial, or not-for-profit sectors.\\

\noindent\textbf{References}
\bibliographystyle{apalike}

\end{document}